\documentclass[aps,twocolumn,preprintnumbers,amsmath,amssymb,superscriptaddress,floatfix,nofootinbib]{revtex4}

\usepackage{lipsum}
\usepackage{graphicx}
\usepackage{epsfig}
\usepackage{epstopdf}
\usepackage{hyperref}
\usepackage{amsmath}
\usepackage{amsfonts}
\usepackage{amssymb}

\usepackage{float}

\begin{document}

\title{$a_0(980)-f_0(980)$ mixing in $\chi_{c1} \to \pi^0 f_0(980) \to \pi^0 \pi^+ \pi^-$ and $\chi_{c1} \to \pi^0 a_0(980) \to \pi^0 \pi^0 \eta$}

\author{M.~Bayar}
\email{melahat.bayar@kocaeli.edu.tr}
\affiliation{Department of Physics, Kocaeli University, 41380, Izmit, Turkey}
\affiliation{Departamento de F\'{\i}sica Te\'orica and IFIC, Centro Mixto Universidad de Valencia-CSIC Institutos de Investigaci\'on de Paterna, Aptdo. 22085, 46071 Valencia, Spain}

\author{V.~R.~Debastiani}
\email{vinicius.rodrigues@ific.uv.es}
\affiliation{Institute of Modern Physics, Chinese Academy of
Sciences, Lanzhou 730000, China}
\affiliation{Departamento de F\'{\i}sica Te\'orica and IFIC, Centro Mixto Universidad de Valencia-CSIC Institutos de Investigaci\'on de Paterna, Aptdo. 22085, 46071 Valencia, Spain}

\date{\today}

\begin{abstract}

We study the isospin breaking in the reactions $\chi_{c1} \to \pi^0 \pi^+ \pi^-$ and $\chi_{c1} \to \pi^0 \pi^0 \eta$ and its relation to the $a_0(980) - f_0(980)$ mixing, which was measured by the BESIII Collaboration. We show that the same theoretical model previously developed to study the $\chi_{c1} \to \eta \pi^+ \pi^-$ reaction (also measured by BESIII), and further explored in the predictions to the $\eta_{c} \to \eta \pi^+ \pi^-$, can be successfully employed in the present study. We assume that the $\chi_{c1}$ behaves as an $SU(3)$ singlet to find the weight in which trios of pseudoscalars are created, followed by the final state interaction of pairs of mesons to describe how the $a_0(980)$ and $f_0(980)$ are dynamically generated, using the chiral unitary approach in coupled channels. The isospin violation is introduced through the use of different masses for the charged and neutral kaons, either in the propagators of pairs of mesons created in the $\chi_{c1}$ decay, or in the propagators inside the $T$ matrix, constructed through the unitarization of the scattering and transition amplitudes of pairs of pseudoscalar mesons. We find that violating isospin inside the $T$ matrix makes the $\pi^0\eta \to \pi^+\pi^-$ amplitude nonzero, which gives an important contribution and also enhances the effect of the $K\bar{K}$ term. We also find that the most important effect in the total amplitude is the isospin breaking inside the $T$ matrix, due to the constructive sum of $\pi^0\eta \to \pi^+\pi^-$ and $K\bar{K} \to \pi^+ \pi^-$, which is essential to get a good agreement with the experimental measurement of the mixing.\\ \vspace{-7pt}

{\fontsize{8}{10}\selectfont \textbf{Keywords:} Isospin-breaking; $a_0(980)-f_0(980)$ mixing; Charmonium decays; Scalar meson states.} 
\end{abstract}

\maketitle

\vspace{-22pt}

\section{Introduction}

\vspace{-5pt}

The nature of the scalar mesons $a_0(980)$ and $f_0(980)$ has been a topic of much discussion since their discovery decades ago. Several models have been proposed, from regular $q\bar{q}$ to more exotic configurations like tetraquarks $qq\bar{q}\bar{q}$, hybrids $q\bar{q}g$ and meson molecules \cite{Jaffe,Weinstein4q,GodfreyRelat,WeinsteinKK,Oller,vanBeveren,HadroMol}. In this context, the isospin-violating mixing of $f_0(980)$ and $a_0^0(980)$ presents an opportunity to filter different proposals and constrain parameters in the models.

In Ref.~\cite{Achasov:1997ih}, the possibility of observing these scalar mesons in the reaction $e^+ e^- \to \gamma \pi^0 \pi^0 (\eta)$ was already discussed along with their different interpretations as $s\bar{s}$ states, tetraquarks or $K\bar{K}$ molecules. Their mixing was first suggested in Ref.~\cite{Achasov:1979xc} and its connection to the difference in the mass of the charged and neutral kaons was already seen as the main source of the isospin symmetry violation. Also, in Ref.~\cite{Krehl:1996rk} the scattering amplitudes of $\pi\pi$ and $\pi\eta$ were studied with the J\"{u}lich meson exchange model and it was found that the cross sections for $\pi\pi \to \pi\eta$ would be nonzero, indicating again the possibility of $a_0(980) - f_0(980)$ mixing.

There are several reactions where this isospin-breaking mixing appears, for instance, in the decay $\eta(1405) \to \pi^0 f_0(980)$ and $\eta(1405) \to \pi^0 a_0(980)$ \cite{BESIII:2012aa}, which was studied in Ref.~\cite{Aceti:2012dj} using the chiral unitary approach. The same puzzle seemed to be present in the decay of the $\eta(1475)$ and both problems were discussed in Refs.~\cite{Wu:2011yx,Wu:2012pg}, where the interesting role of the triangular singularities in enhancing the isospin violation was shown. This reaction was also discussed in Ref.~\cite{Achasov:2015uua}, and in the decay $f_1(1285) \to \pi^0 f_0(980)$ was studied in the same lines.

Recently, another case where a triangular singularity reinforces the isospin breaking in the $a_0(980) - f_0(980)$ mixing was studied in Ref.~\cite{Sakai:2017iqs}, indicating that the reaction $D_s^+ \to \pi^+ \pi^0 a_0(980) (f_0(980))$ could bring further information on this subject.

Also recently, the role of the $a_0(980) - f_0(980)$ mixing was investigated in the $D^0 \to K^0_S \,\pi^+ \pi^-$ and $D^0 \to K^0_S \, \eta \pi^0$ decays \cite{Achasov:2017zhu}; and also in the $D^+_s\to\eta\pi^0\pi^+$ decay \cite{Achasov:2017edm}, showing new possible reactions to investigate this topic. In Ref.~\cite{Wang:2016wpc} several possibilities of $D_s$ and $B_s$ decays have been proposed and it is argued that the $a_0(980) - f_0(980)$ mixing could be experimentally determined with high precision.

One of the first attempts to quantitatively relate the $a_0(980) - f_0(980)$ mixing with experimental data was made in Ref.~\cite{Close:2000ah}, through the analysis of an enhancement in the production rate of the $a_0(980)$ relative to the $a_2(1320)$ in $pp \to p_s (\eta \pi^0) p_f$. However, questions about other secondary effects related to $G$-parity were raised in Ref.~\cite{Wu:2007jh}, which could affect the assumptions made in Ref.~\cite{Close:2000ah}.

The mixing of these scalar mesons in the radiative $\phi$ decay was discussed in Refs.~\cite{Close:2001ay,Achasov:2002hg}, while the photoproduction of $f_0(980)$ and $a_0(980)$ was studied in Ref.~\cite{Kerbikov:2000pu}, with emphasis on the isospin-violating mixing due to the mass difference of kaons and the role of polarized photons and protons.

The decay of $\eta'$ has also been a topic where this mixing was investigated. For instance, in Ref.~~\cite{Donskov:2009ri} the reaction $\eta' \to \eta\pi^0\pi^0$ was studied in the framework of the isobar model, where $f_0(500)$ was also included in the analysis. Similarly, the decays $\eta' \to 3\pi^0$ and $\eta' \to \pi^0\pi^+\pi^-$ were considered in Ref.~\cite{Likhoded:2010yv}, both recently measured by the BESIII Collaboration \cite{Ablikim:2016frj}. After this measurement, the decay $\eta \to 3\pi$ was studied in Ref.~\cite{Albaladejo:2017hhj} with an extended chiral Khuri–Treiman formalism, where the $a_0(980)$ and $f_0(980)$ are taken into account in the dispersive integrals.

Other reactions have been proposed to search for the $f_0(980)$ and $a_0(980)$ mixing, like the $p\, n \to\, d \, a_0$ in Ref.~\cite{Kudryavtsev:2001ee}. This reaction was also studied in Ref.~\cite{Grishina:2001zj}, where two more reactions were proposed: the $p\, d \to\, ^3\mathrm{He}\,/\, ^3\mathrm{H} \,a_0$ and the $d\, d \to\, ^4\mathrm{He} \,a_0$. Also, in Ref.~\cite{Achasov:2003se} it was suggested performing polarized target experiments on the reaction $\pi^-\,p \to \eta \pi^0 n$, where the mixing would be detected through the presence of a jump in the azimuthal asymmetry in the $\pi^0 \eta$ $S$-wave production cross section around the $K\bar{K}$ threshold.

Searching for a reaction where the isospin breaking could be measured unambiguously, the decay $J/\psi \to \phi f_0(980) \to \phi a_0(980) \to \phi \pi^0 \eta$ was proposed in Ref.~\cite{Wu:2007jh}, where it was assumed that first there would be the formation of the $f_0(980)$, which then would make a transition to $a_0(980)$ violating isospin conservation and finally the later would decay into $\pi^0 \eta$. The background of other reactions was analysed and the conclusion was that one should expect a narrow peak in the $\pi^0 \eta$ invariant mass with a width of about 8 MeV in the region of the $K\bar{K}$ threshold, which would come from the difference in the mass of the charged and neutral kaons, and would be clearly distinguishable from the broad structure of other background process.

The reaction $J/\psi \to  \phi \pi^0 \eta$ was also investigated in Ref.~\cite{Hanhart:2007bd}, where the chiral unitary approach was used to study the $a_0(980) - f_0(980)$ mixing, considering the difference in quark masses and also one-photon exchange between charged mesons. It was shown that this mixing was indeed the most important isospin-breaking effect and could be extracted from experiment through that reaction.

Next, the question whether there would be a difference in the inverse isospin-breaking process, where the $a_0(980)$ would make a transition to the $f_0(980)$, the complementary reaction $\chi_{c1} \to \pi^0 a_0(980) \to \pi^0 f_0(980) \to \pi^0 \pi^+ \pi^-$ was proposed in Ref.~\cite{Wu:2008hx}, and it was found that one could indeed expect different rates of mixing. The uncertainty of these calculations were attributed essentially to the different parameters extracted from different theoretical models or experimental measurements of these two scalar mesons.

Some time later, the two reactions proposed in Refs.~\cite{Wu:2007jh,Hanhart:2007bd,Wu:2008hx} were measured by the BESIII Collaboration \cite{Ablikim:2010aa}, the isospin-forbidden production of $a_0(980)$ in the decay $J/\psi \to \phi \pi^0 \eta$ and the isospin-forbidden production of $f_0(980)$ in the decay $\chi_{c1} \to \pi^0 \pi^+ \pi^-$. The mixing in both reactions was determined through the fraction of the branching ratios with their corresponding isospin-allowed process \cite{Ablikim:2010aa}, respectively the $J/\psi \to \phi \pi^+ \pi^-$ (where the $f_0(980)$ shows up) measured by the BES Collaboration \cite{Ablikim:2004wn}, and the $\chi_{c1} \to \pi^0 \pi^0 \eta$ (where the $a_0^0(980)$ shows up). As argued in Ref.~\cite{Wu:2008hx}, the later reaction could be compared to the $\chi_{c1} \to \eta \pi^+ \pi^-$ (where the $a_0^\pm(980)$ shows up clearly), since by isospin symmetry the same production rate is expected for $\chi_{c1} \to \pi^0 a_0^0(980)$ as in $\chi_{c1} \to \pi^\pm a_0^\mp(980)$ (with $a_0^\mp(980)\to \eta\pi^\mp$). The $\chi_{c1} \to \eta \pi^+ \pi^-$ was measured by the CLOE Collaboration \cite{Athar:2006gq} and recently by BESIII \cite{Kornicer:2016axs} with high statistics.

After the BESIII experiment \cite{Ablikim:2010aa}, the reaction $J/\psi \to \phi \pi^0 \eta$ was studied in Ref.~\cite{Roca:2012cv} using the chiral unitary approach, where the importance of other mechanisms was also shown, like the sequential exchange of vector and axial-vector mesons to obtain a good agreement with the data. Also based on this experiment, a study of the amount of $K\bar{K}$ in the $a_0(980)$ and $f_0(980)$ was developed in Ref.~\cite{Sekihara:2014qxa} using the chiral unitary approach and the Flatt\'{e} parametrization, where the mixing of these scalar mesons, formulated in a similar manner of Refs.~\cite{Wu:2007jh,Wu:2008hx}, was used to constrain their parameters and compositeness.

Not much theoretical work has been done to describe the other isospin-breaking reaction also measured by BESIII, the $\chi_{c1} \to \pi^0 \pi^+ \pi^-$, in which we focus.
There is one reason to tackle this reaction at this stage, since the recent experiment by BESIII \cite{Kornicer:2016axs} on the $\chi_{c1} \to \eta \pi^+ \pi^-$ reaction has brought new light into this problem. Indeed, the process was studied theoretically in Ref.~\cite{Liang:2016hmr}, with the basic assumption that the $\chi_{c1}$ is an $SU(3)$ singlet due to its $c\bar{c}$ structure. The different $SU(3)$ scalar structures with three mesons were discussed in Ref.~\cite{Debastiani:2016ayp} in the study of the $\eta_c \to \eta \pi^+ \pi^-$ reaction \cite{Debastiani:2016ayp,Debastiani:2017jcr}, supporting the structure used in Ref.~\cite{Liang:2016hmr} by means of which a good agreement with the experimental data of $\chi_{c1} \to \eta \pi^+ \pi^-$ \cite{Kornicer:2016axs} was found. This information is important for the $\chi_{c1} \to \pi^0 \pi^+ \pi^-$ reaction since it provides the weights of different trios of pseudoscalar mesons that can be formed, prior to their final state interaction from where the $f_0(980)$ and $a_0(980)$ resonances emerge. The use of this information and of the chiral unitary approach to deal with the interaction of pairs of pseudoscalars allows a thorough investigation of this process, clarifying the mechanisms that lead to isospin breaking, and providing for the first time a quantitative description of the $f_0(980)$ and $a_0(980)$ production with a ratio of strengths in agreement with the BESIII \cite{Ablikim:2010aa} experimental data.

\vspace{-15pt}

\section{Formalism}

\vspace{-5pt}

We follow a similar approach to the one of Refs.~\cite{Liang:2016hmr,Debastiani:2016ayp} in order to study the $\chi_{c1} \to \pi^0 f_0(980) \to \pi^0 \pi^+ \pi^-$ and $\chi_{c1} \to \pi^0 a_0(980) \to \pi^0 \pi^0 \eta$ decays. We assume that the $\chi_{c1}$ behaves as a flavor $SU(3)$ singlet since it is essentially a $c \bar{c}$ state. Hence we use the following $\phi$ matrix, with $ \eta - \eta' $ mixing, to construct an $SU(3)$ singlet with trios of pseudoscalar mesons:
\begin{widetext}
\begin{equation}\label{eq:phi}
\phi \equiv \left(
           \begin{array}{ccc}
             \frac{1}{\sqrt{2}}\pi^0 + \frac{1}{\sqrt{3}}\eta + \frac{1}{\sqrt{6}}\eta' & \pi^+ & K^+ \\
             \pi^- & -\frac{1}{\sqrt{2}}\pi^0 + \frac{1}{\sqrt{3}}\eta + \frac{1}{\sqrt{6}}\eta' & K^0 \\
            K^- & \bar{K}^0 & -\frac{1}{\sqrt{3}}\eta + \sqrt{\frac{2}{3}}\eta' \\
           \end{array}
         \right).
\end{equation}
\end{widetext}

There are three independent $SU(3)$ scalars from $\phi\phi\phi$: ${\rm Trace}(\phi\phi\phi)$, ${\rm Trace}(\phi){\rm Trace}(\phi\phi)$ and $[{\rm Trace}(\phi)]^{3}$. However, in Refs.~\cite{Debastiani:2016ayp,Debastiani:2017jcr}, the authors discuss these three $SU(3)$ scalars and conclude that only the structure ${\rm Trace}(\phi \phi \phi)$ yields results in good agrement with the recent experiment of BESIII \cite{Kornicer:2016axs} on the $\chi_{c1} \to \eta \pi^+ \pi^-$ decay. In fact, that is expected from large $N_c$ counting, since each time one takes a trace a factor $1/N_c$ is introduced \cite{Manohar:1998xv,Guo:2013nja}. Besides, if one does not include the $\eta_1$ (which we do through the inclusion of $ \eta - \eta' $ mixing, in order to relate the $\phi$ matrix with the $q\bar{q}$ matrix \cite{Liang:2016hmr}) but instead take $\eta\to\eta_8$ and no $\eta'$, then ${\rm Trace}(\phi)=0$.

Therefore, in the present work we also adopt ${\rm Trace}(\phi\phi\phi)$ as the $SU(3)$ singlet to investigate the $\chi_{c1} \to \pi^0 \pi^+ \pi^-$ and $\chi_{c1} \to \pi^0 \pi^0 \eta$ decays.

\begin{figure}[H]
 \centering
 \includegraphics[width=0.46\textwidth]{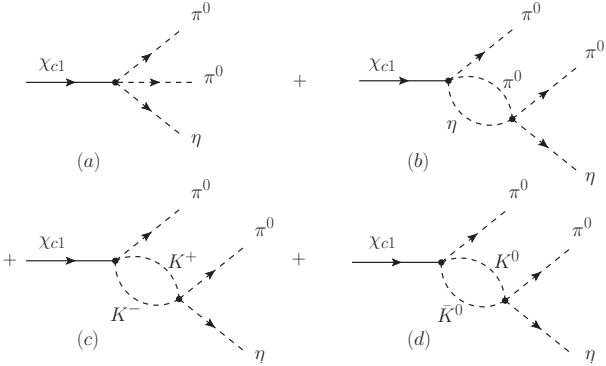}
 \caption{Diagrams involved in the $a_0(980)$ production in the $\chi_{c1} \to \pi^0 a_0(980) \to \pi^0 \pi^0 \eta$ reaction: (a) tree-level; and rescattering of (b) $\pi^0 \eta$, (c) $K^+ K^-$, (d) $K^0 \bar{K}^0$.} \label{a0diagrams}
\end{figure}

\begin{figure}[H]
 \centering
 \includegraphics[width=0.45\textwidth]{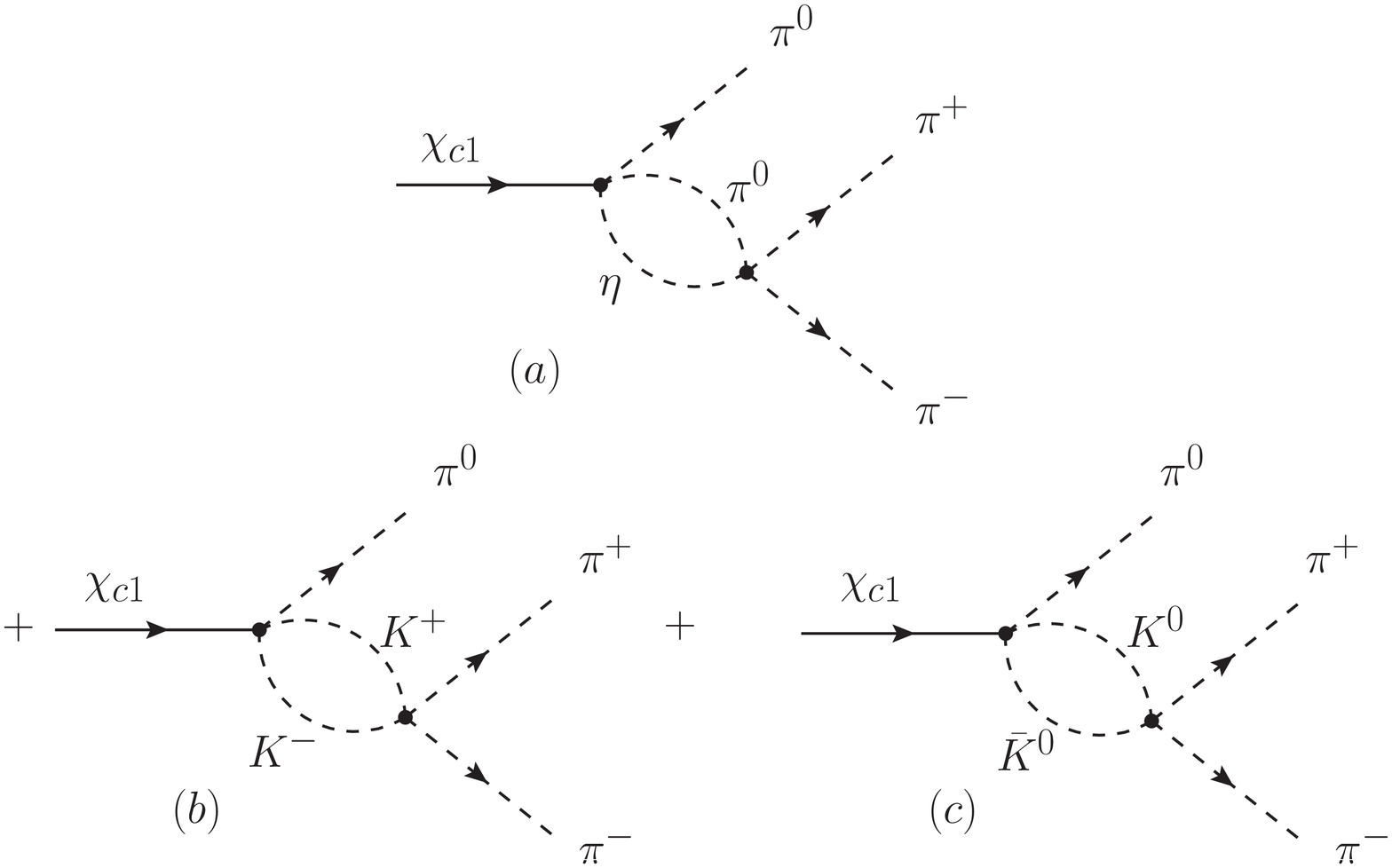}
 \caption{Diagrams involved in the $f_0(980)$ production in the $\chi_{c1} \to \pi^0 f_0(980) \to \pi^0 \pi^+ \pi^-$ reaction: rescattering of (a) $\pi^0 \eta$, (b) $K^+ K^-$, (c) $K^0 \bar{K}^0$.} \label{f0diagrams}
\end{figure}

Then we perform the trace of $\phi\phi\phi$ using the matrix of pseudoscalar mesons in Eq. (\ref{eq:phi}) and select only the terms that have at least one $\pi^{0}$. Thus we obtain the combinations $\pi^{0} \pi^{0} \eta$ and $ \pi^{0} K \bar{K}$, as follows
\begin{equation}
 {\rm Trace}(\phi\phi\phi) =    \sqrt{3}\,\pi^{0}\pi^{0}\eta
 +\dfrac{\pi^{0}}{\sqrt{2}} (3\,K^{+} K^{-} - 3\,K^0 \bar{K}^{0}),
\label{Eq:Trace}
\end{equation}
where we have neglected the $\eta'$ components which we omit in the coupled channels because of its large mass and small couplings to these scalar mesons. Then Eq. \eqref{Eq:Trace} tells us the weight by which trios of pseudoscalars are produced in the first step of the $\chi_{c1}$ decay. The next step consists of letting these mesons interact in coupled channels such that the final $\pi^{0} \pi^{0} \eta$ or $\pi^{0} \pi^{+} \pi^{-}$ mesons are produced.
The  diagrams of $a_0(980)$ production in the $\chi_{c1} \to \pi^0 a_0(980) \to \pi^0 \pi^0 \eta$ reaction are shown in Fig. \ref{a0diagrams} and for $f_0(980)$ production in the $\chi_{c1} \to \pi^0 f_0(980) \to \pi^0 \pi^+ \pi^-$ reaction in Fig. \ref{f0diagrams}.

The quantum numbers of the $\chi_{c1}$ are $I^{G}(J^{PC}) = 0^{+}(1^{++})$,
while the quantum numbers of the $a_0(980)$ and $f_0(980)$ are $1^{-}(0^{++})$ and $0^{+}(0^{++})$, respectively. If the $\pi^{0} \eta$ and $\pi^{+} \pi^{-}$ are in $S$-wave to create the $a_0(980)$ and $f_0(980)$, the remaining $\pi^{0}$ must be in $P$-wave to conserve angular momentum and parity.

Following Refs.~\cite{Debastiani:2016ayp,Liang:2016hmr}, for $a_0(980)$ ($\pi^{0}\eta$) production we have a structure at tree-level like
\begin{equation}\label{eq:tree}
t = V_{p} ~ \vec{\epsilon}_{\chi_{c1}} \cdot \vec{p}_{\pi^0 },
\end{equation}
where the factor $V_{p}$ is a constant coefficient related to the basic dynamics of $\chi_{c1} \rightarrow$ three mesons. It is taken as a global factor that can be adjusted to the data. For comparison purposes we take the same value that was used in Ref.~\cite{Liang:2016hmr}.

Then the full amplitude for the isospin-allowed $a_0(980)$ production (with final state $\pi^0\eta$) is obtained considering also the rescattering of the pairs of mesons as indicated in Fig. \ref{a0diagrams},
\begin{equation}\label{eq:tfull}
t =   \vec{\epsilon}_{\chi_{c1}} \cdot \vec{p}_{\pi^0 } ~ \tilde{t}_{ \pi^{0}  \eta},
\end{equation}
with
\begin{eqnarray}
\tilde{t}_{\pi^{0}\eta} &=&  V_{p}~(h_{\pi^{0}\eta} + h_{\pi^{0}\eta}~G_{\pi^{0}\eta}~t_{\pi^{0}\eta \to \pi^{0}\eta} \nonumber \\
&+& h_{K^{+}K^{-}}~G_{K^{+} K^{-}}~t_{K^{+}K^{-} \to \pi^{0}\eta} \nonumber \\
&+& h_{K^{0}\bar{K}^{0}}~G_{K^{0}\bar{K}^{0}}~t_{K^{0}\bar{K}^{0} \to \pi^{0}\eta}) \, ,
\label{Eq:tpeta}
\end{eqnarray}
where the weights $h_{i}$ are obtained from Eq. \eqref{Eq:Trace}: $h_{\pi^{0}\eta} = 2\sqrt{3}$, $h_{K^{+}K^{-}} = 3 / \sqrt{2}$ and $h_{K^{0}\bar{K}^{0}} = -3 / \sqrt{2}$.
Note that $h_{\pi^{0}\eta}$ has an extra factor 2 with respect to the coefficient $\sqrt{3}$ for the $\pi^0\pi^0\eta$ field in Eq.~\eqref{Eq:Trace}, since with the production of two $\pi^0$ we will have the terms $\partial_i\pi^0\,\pi^0 + \pi^0\,\partial_i\pi^0$. The functions $G_{ij}$ are the same used in the on-shell factorization of the Bethe-Salpeter equation $T=(1-VG)^{-1}V$ to account for all the meson-meson loops \cite{Oller}
\begin{equation}
G_l=i\int\frac{d^4q}{(2\pi)^4}\frac{1}{q^2-m_1^2+i\epsilon}\frac{1}{(P-q)^2-m_2^2+i\epsilon}\ ,
\label{eq:loopex}
\end{equation}
where $m_1$ and $m_2$ are the masses of the two meson of the $l$-channel, which in charge basis are: 1) $\pi^+\pi^-$, 2) $\pi^0\pi^0$, 3) $K^+K^-$, 4) $K^0\bar K^0$, 5) $\eta\eta$ and 6) $\pi^0\eta$. The loop functions are then regularized with a cutoff $q_{max}\sim600$ MeV, the same used in Refs.~\cite{Debastiani:2016ayp,Liang:2016hmr}. After the integration in $q^0$ and $\cos\theta$ we have
\begin{align}
G&=\int_0^{q_{max}}\frac{\mathbf{q}^2dq}{(2\pi)^2}\frac{\omega_1 + \omega_2}{\omega_1\omega_2[(P^0)^2-(\omega_1 + \omega_2)+i\epsilon]}\ ,\\
\nonumber\omega_i &= \sqrt{\mathbf{q}^2+m_i^2}, \quad (P^0)^2 = s.
\label{eq:loopex}
\end{align}

This approach is in the same line of Ref.~\cite{Roca:2012cv}, but it is different from the approach of Ref.~\cite{Wu:2008hx}, where it was assumed that the isospin-forbidden production of $f_0(980)$ comes from a transition $a_0(980) \to f_0(980)$, related to the phase space available in the propagators of pairs of mesons. On the other hand, we assume that the $f_0(980)$ emerges from the $\chi_{c1}$ decay, stemming from the meson-meson loops, without going first through the $a_0(980)$ production.

The difference between the $K^+K^-$ and $K^0\bar K^0$ loops is convergent, and useful to investigate the $f_0(980)$ production, but in order to deal with the $a_0(980)$ production and study the whole problem quantitatively, one must face the divergent behaviour of all the propagators. For that we have a simple solution of employing the same cutoff used to regularize the loops inside the $T$ matrix, which yields results in good agreement with the $\chi_{c1} \to \eta \pi^+ \pi^-$ decay, as shown in Ref.~\cite{Liang:2016hmr}.

Now for the isospin-forbidden $f_0(980)$ production (with final state $\pi^{+}\pi^{-}$) we have no tree-level, and we consider only the rescattering diagrams, as indicated in Fig. \ref{f0diagrams},
\begin{equation}\label{eq:tfullpp}
t =  \vec{\epsilon}_{\chi_{c1}} \cdot \vec{p}_{\pi^0 } ~ \tilde{t}_{\pi^{+}\pi^{-}},
\end{equation}
where
\begin{eqnarray}
\tilde{t}_{\pi^{+}\pi^{-}} &=&  V_{p}~(h_{\pi^{0}\eta}~G_{\pi^{0}\eta}~t_{\pi^{0}\eta \to \pi^{+}\pi^{-}} \nonumber \\
&+& h_{K^{+}K^{-}}~G_{K^{+}K^{-}}~t_{K^{+}K^{-} \to \pi^{+}\pi^{-}} \nonumber \\
&+& h_{K^{0}\bar{K}^{0}}~G_{K^{0}\bar{K}^{0}}~t_{K^{0}\bar{K}^{0} \to \pi^{+}\pi^{-}}) \, .
\label{Eq:tpp}
\end{eqnarray}

Note that if we consider isospin symmetry this amplitude would be identically zero, because $\pi^0\eta \to \pi^+ \pi^-$ would not conserve isospin $-$ since we consider $\pi^+ \pi^-$ in $I=0$ to create the $f_0(980)$ $-$ and the terms with kaons would cancel due to the minus sign in $h_{K^{0}\bar{K}^{0}}$. Indeed, we can interpret the last two terms as $K\bar{K}$ in isospin 1 basis, which again, would not go to $\pi^+ \pi^-$ in $I=0$.

Therefore, the only way to have $f_0(980)$ production is by introducing isospin breaking, which we do with the use of the different masses for the charged and neutral kaons. We introduce isospin violation from two sources, one comes from $G_{K^{+} K^{-}}$ and $G_{K^{0}\bar{K}^{0}}$, the first loops of rescattering with $K^{+} K^{-}$ and $K^{0}\bar{K}^{0}$ pairs in Eqs. \eqref{Eq:tpeta} and \eqref{Eq:tpp}. In the case of $\pi^{0} \eta$ production the $K^{+} K^{-}$ and $K^{0} \bar{K}^{0}$ terms add, but in the case of the $\pi^{+} \pi^{-}$ production they subtract, and would cancel if the masses of the kaons were equal, but not when they are different.

The other source comes from the $T$ matrix that we construct with different kaon masses in the propagators inside the Bethe-Salpeter equation $T=(1-VG)^{-1}V$ \cite{Oller}, which we use to obtain the scattering and transition amplitudes $t_{i \to \pi^0\eta}$ and $t_{i \to \pi^+\pi^-}$. This way, the $t_{\pi^{0}\eta \to \pi^{+}\pi^{-}}$ transition will also be nonzero $-$ because of the coupled channels interaction $-$ when we introduce isospin breaking inside the $T$ matrix.

Finally, for the case of $ a_0(980)$ production we can write the invariant mass distribution as
\begin{equation}
\dfrac{d \Gamma}{dM_{\rm inv}(\pi^{0}  \eta)} = \dfrac{1}{(2 \pi)^{3}} \dfrac{1}{4 M^{2}_{\chi_{c1} }}\frac{1}{3} p^{2}_{\pi^{0}} ~p_{\pi^{0}} ~\tilde{p}_{\eta} \vert \tilde{t}_{\pi^{0}  \eta} \vert^{2},
\label{Eq:dp0E}
\end{equation}
where
\begin{equation}
p_{\pi^{0}} = \dfrac{\lambda^{1/2}(M^{2}_{\chi_{c1}}, m^{2}_{\pi^{0}},M^{2}_{\rm inv}(\pi^{0}  \eta) )}{2 M_{\chi_{c1}} },
\label{Eq:Pp0}
\end{equation}
and
\begin{equation}
\tilde{p}_{\eta} = \dfrac{\lambda^{1/2}(M^{2}_{\rm inv}(\pi^{0}  \eta), m^{2}_{\pi^{0}},m^{2}_{\eta})}{2 M_{\rm inv}(\pi^{0}  \eta) }.
\label{Eq:Peta}
\end{equation}

On the other hand, for the case of $f_0(980)$ production, the invariant mass distribution reads
\begin{equation}
\dfrac{d \Gamma}{dM_{\rm inv}( \pi^{+}  \pi^{-})} = \dfrac{1}{(2 \pi)^{3}} \dfrac{1}{4 M^{2}_{\chi_{c1} }}\frac{1}{3} p^{2}_{\pi^{0}} ~p_{\pi^{0}} ~\tilde{p}_{\pi^{+}} \vert \tilde{t}_{\pi^{+}  \pi^{-}} \vert^{2},
\label{Eq:dpppm}
\end{equation}
with
\begin{equation}
p_{\pi^{0}} = \dfrac{\lambda^{1/2}(M^{2}_{\chi_{c1}}, m^{2}_{\pi^{0}},M^{2}_{\rm inv}(\pi^{+}  \pi^{-}) )}{2 M_{\chi_{c1}} },
\label{Eq:Pp0}
\end{equation}
and
\begin{equation}
\tilde{p}_{\pi^{+}} = \dfrac{\lambda^{1/2}(M^{2}_{\rm inv}(\pi^{+}  \pi^{-}), m^{2}_{\pi^{+}},m^{2}_{\pi^{-}})}{2 M_{\rm inv}(\pi^{+}  \pi^{-}) }.
\label{Eq:Peta}
\end{equation}

\section{Results}

First we notice that all our results are calculated using average pion masses, but the effect of using different pion masses, either inside the $T$ matrix and/or in the external propagator $G_{\pi^0\eta}$ and in Eqs. \eqref{Eq:dp0E} to \eqref{Eq:Peta} is completely negligible (as we have checked) in comparison with the really important effect of using different masses for the charged and neutral kaons, as expected.

\begin{figure}[H]
 \centering
 \includegraphics[width=0.45\textwidth]{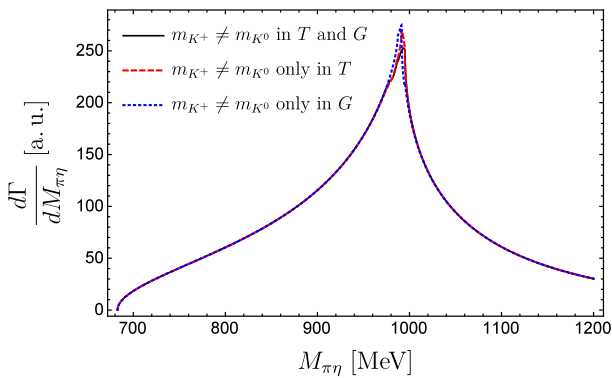}
 \caption{Invariant mass distribution of $\pi^0\eta$ in the $\chi_{c1} \to \pi^0 a_0(980) \to \pi^0 \pi^0 \eta$ reaction. (See text for explanations).} \label{a0}
\end{figure}

\begin{figure}[H]
 \centering
 \includegraphics[width=0.45\textwidth]{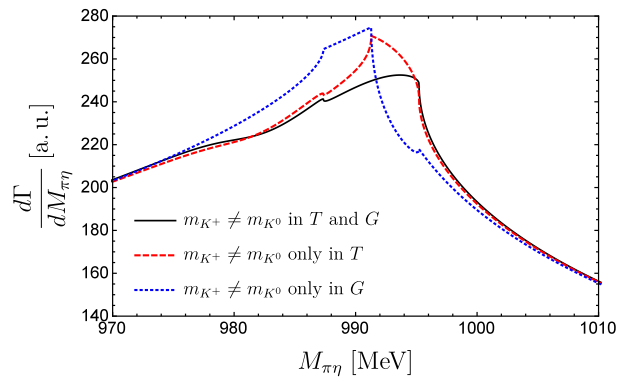}
 \caption{Zoom around the $a_0(980)$ peak in the invariant mass distribution of $\pi^0\eta$ in the $\chi_{c1} \to \pi^0 a_0(980) \to \pi^0 \pi^0 \eta$ reaction. (See text for explanations).} \label{a0zoom}
\end{figure}

We show in Figs.~\ref{a0} and \ref{a0zoom} the invariant mass distribution $d\Gamma/dM_{\pi\eta}$ from Eq. \eqref{Eq:dp0E}, where the shape of the $a_0(980)$ is clear. The solid line represents the case where different masses for the charged and neutral kaons are used in the propagators inside the $T$ matrix and also in the first rescattering loops $G_{K^{+} K^{-}}$ and $G_{K^{0}\bar{K}^{0}}$, as discussed in the formalism. The dashed line is the case where the different masses are used only inside the $T$ matrix and the dotted line only in the first rescattering loops $G_{K^{+} K^{-}}$ and $G_{K^{0}\bar{K}^{0}}$. As we can see, there is only a small difference in the curves around the $K\bar{K}$ threshold. By looking closer into this region, one can see in Fig.~\ref{a0zoom} that in the three curves there is a small cusp effect in $M_{\pi\eta}$ at $2\,m_{K^+}$ and $2\,m_{K^0}$; and in the dashed and dotted line, where the isospin-average kaon mass $\langle m_K \rangle$ is also used, the $a_0(980)$ peak appears at $2\,\langle m_K \rangle$.

Notice that there is an interesting comparison to be made with the reaction $\chi_{c1} \to \eta \pi^+ \pi^-$. In this case, the $\chi_{c1} \to \pi^0 \pi^0 \eta$ has the same isospin content, and we can see for instance in Eq.~(7) of Ref.~\cite{Liang:2016hmr} that when the $\pi^+$ is in $P$-wave the $\pi^-\eta$ term has the same weight as the $\pi^0\eta$ in Eq.~\eqref{Eq:Trace} $-$ after the inclusion of the statistical factor $-$ while the $K^0K^-$ term has the same weight as the $(K^{+} K^{-} - \,K^0 \bar{K}^{0})/\sqrt{2}$, and both are $K\bar{K}$ in isospin 1. The same is valid for Eq.~(8) of Ref.~\cite{Liang:2016hmr}, when the $\pi^-$ is in $P$-wave. Indeed, if we look at Fig.~6 of  Ref.~\cite{Liang:2016hmr}, we see that the intensity of the $a_0(980)$ peak is exactly twice as here, in Figs.~\ref{a0} and \ref{a0zoom}, since there we had the sum of both contributions of $\pi^+\eta$ and $\pi^-\eta$, and here we have only $\pi^0\eta$ (notice the extra factor 2 in Eq.~(24) of that reference).
\begin{figure}[H]
\centering
\includegraphics[width=0.45\textwidth]{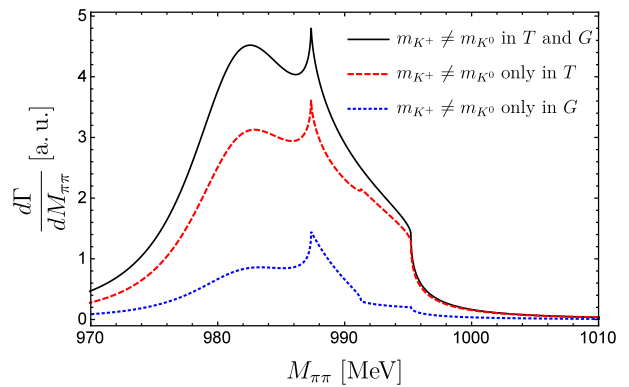}
\caption{Invariant mass distribution of $\pi^+\pi^-$ in the $\chi_{c1} \to \pi^0 f_0(980) \to \pi^0 \pi^+ \pi^-$ reaction. (See text for explanations).} \label{f0}
\end{figure}

For the isospin-breaking production of $f_0(980)$, we show in Fig.~\ref{f0} the invariant mass distribution $d\Gamma/dM_{\pi\pi}$ from Eq. \eqref{Eq:dpppm}. We have a narrow peak around the threshold of $K\bar{K}$ similar to the results of the literature \cite{Wu:2008hx}. We can see clearly the effect of the two different thresholds, at $M_{\pi\pi}$ equal to $2\,m_{K^+}$ and $2\,m_{K^0}$, and for the case of the dashed and dotted lines, we also see the cusp effect at $M_{\pi\pi}$ equal to $2\,\langle m_K \rangle$.

The bump around 980 MeV of $d\Gamma / dM_{\pi\pi}$ is a manifestation of the ``good'' $f_0(980)$, that can be obtained if one considers the sum of the amplitudes $t_{K^{+}K^{-} \to \pi^{+}\pi^{-}}$ and $t_{K^{0}\bar{K}^{0} \to \pi^{+}\pi^{-}}$, with a plus sign, which would be the scattering of $K\bar{K} \to \pi\pi$ in isospin 0. In this work, as in Ref.~\cite{Liang:2016hmr}, we use a cutoff of 600 MeV to regularize the loops inside and outside the $T$ matrix, which better fits the data and yields the peak position of the $f_0(980)$ in $K\bar{K} \to \pi\pi$ amplitude at 980 MeV, in agreement with the majority of experimental measurements. This bump can be translated in the direction of the $K^{+}K^{-}$ threshold by lowering the cutoff to about 560 MeV, making it less bound and the curve more similar to results of Ref.~\cite{Wu:2008hx}, for instance. However,
the use of 600 MeV is more appropriate since it is contrasted with the $f_0(980)$ production in isospin-allowed experiments. The shape of Fig.~\ref{f0} also tells us that a very precise measurement of the $\pi^{+}\pi^{-}$ invariant mass distribution in the $\chi_{c1} \to \pi^0 \pi^+ \pi^-$ reaction could help to constrain the model and determine the $f_0(980)$ and $a_0(980)$ parameters precisely.

Another important point to notice is that, according to our findings, the isospin breaking inside the $T$ matrix turns out to be more important than from the external kaon loops $G_{K^{+} K^{-}}$ and $G_{K^{0}\bar{K}^{0}}$, as shown in Fig. \ref{f0}. This seems to go against what one would naturally expect: that regarding the $K\bar{K}$ interaction, the contribution from the external loops would be more significative than the one from the loops inside the $T$ matrix.
Actually, this is implicitly assumed in Refs.~\cite{Wu:2008hx,Sekihara:2014qxa}, since isospin symmetry is assumed for the coupling of the $f_0(980)$ to the $K^{+} K^{-}$ and $K^{0}\bar{K}^{0}$ components.

To investigate this interesting feature, we look back to Eq.~\eqref{Eq:Trace}, where we see that the contribution coming from $\pi^{0}\eta \to \pi^{+}\pi^{-}$ is:
\begin{align}\label{pi0eta}
   \nonumber  \sqrt{3}\,\pi^{0}\pi^{0}\eta
  &\,\Rightarrow\,
  h_{\pi^{0}\eta}\,~G_{\pi^{0}\eta}~t_{\pi^{0}\eta \to \pi^+\pi^-} \\
  &\;=
  2\sqrt{3}~G_{\pi^{0}\eta}~t_{\pi^{0}\eta \to \pi^+\pi^-},
\end{align}

while the contribution coming from $K\bar{K} \to \pi^+\pi^-$ is:
\begin{align}\label{KKbar}
 \nonumber 3\,\pi^{0}\dfrac{(\,K^{+} K^{-} - \,K^0 \bar{K}^{0})}{\sqrt{2}}
  \,\Rightarrow\,
  & h_{K^{+}K^{-}}~G_{K^{+} K^{-}}~t_{K^{+}K^{-} \to \pi^+\pi^-}\\
\nonumber
+& \, h_{K^{0}\bar{K}^{0}}~G_{K^{0}\bar{K}^{0}}~t_{K^{0}\bar{K}^{0} \to \pi^+\pi^-}\\
 \nonumber
  =&\, \dfrac{3}{\sqrt{2}}G_{K^{+} K^{-}}~t_{K^{+}K^{-} \to \pi^+\pi^-}\\
  -&\, \dfrac{3}{\sqrt{2}}G_{K^{0}\bar{K}^{0}}~t_{K^{0}\bar{K}^{0} \to \pi^+\pi^-}.
\end{align}

\begin{figure}[H]
\centering
\includegraphics[width=0.45\textwidth]{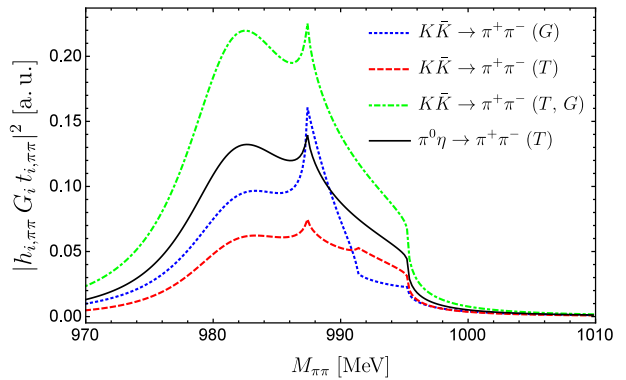}
\caption{Comparison between isolated contribution of $\pi^{0}\eta \to \pi^{+}\pi^{-}$ and $K\bar{K} \to \pi^+\pi^-$. (See text for explanations).} \label{f0pi0etaKKbar}
\end{figure}

Then, we compare in Fig.~\ref{f0pi0etaKKbar} the amplitude square of these two terms. We can see that the $K\bar{K}$ contribution coming from the isospin breaking in $G_{K^{+} K^{-}}$ and $G_{K^{0}\bar{K}^{0}}$ (blue dotted curve), is indeed greater than the one coming from the isospin breaking inside the $T$ matrix (red dashed curve), and the effect is maximized when isospin symmetry is broken in both (green dash-dotted curve).
What happens is that when we violate isospin inside the $T$ matrix, the transition amplitude $t_{\pi^{0}\eta \to \pi^{+}\pi^{-}}$ becomes nonzero (black solid line), which is even bigger than the contribution of $K\bar{K}$ with isospin breaking only inside the $T$ matrix (red dashed curve), and their combined effect turns out to be greater than the isolated effect from $G_{K^{+} K^{-}}$ and $G_{K^{0}\bar{K}^{0}}$ (blue dotted curve).

Therefore, we can conclude here that the $K\bar{K}$ contribution is still the dominant term when isospin symmetry is broken both in $T$ and $G$. However, the effect of isospin-breaking inside the $T$ matrix is of great importance, not just because of the enhancement in the $K\bar{K}$ contribution, but mainly due to the coupled channels interaction that allows the isospin-forbidden $\pi^{0}\eta \to \pi^{+}\pi^{-}$ transition. This is a novel result which is usually neglected in most of the approaches in the topic of the $a_0(980) - f_0(980)$ mixing.

Finally, we show in Table \ref{table} the results of the $a_0(980) - f_0(980)$ mixing in the $\chi_{c1} \to \pi^0 \pi^+ \pi^-$ and $\chi_{c1} \to \pi^0 \pi^0 \eta$ reactions. We calculate it in analogy to Ref.~\cite{Ablikim:2010aa} (where it is taken as the ratio between the branching ratios of the former to the later reaction) by integrating the invariant mass distribution $d\Gamma/dM_{\pi\pi}$ and dividing it by the integrated $d\Gamma/dM_{\pi\eta}$, where the later we calculate in two ways: first we integrate it in the whole mass distribution of the $M_{\pi\eta}$, from $m_\pi + m_\eta$ up to 1200 MeV (around the limit of validity the model) and in the more appropriate range of the $a_0(980)$ resonance, from 885 MeV to 1085 MeV, as done by the BESIII Collaboration in Ref.~\cite{Ablikim:2006vm} (Section IV.C.2).
\begin{center}
\begin{table}[H]
\caption{\label{table} Comparison between experiment and theoretical results for the $a_0(980) - f_0(980)$ mixing in the $\chi_{c1} \to \pi^0 \pi^+ \pi^-$ and $\chi_{c1} \to \pi^0 \pi^0 \eta$ reactions.}
\begin{tabular}{|c|c|}
  \hline
  & $\Gamma(\chi_{c1} \to \pi^0 \pi^+ \pi^-)\,/\,\Gamma(\chi_{c1} \to \pi^0 \pi^0 \eta)$ \\
  \hline
  BESIII \cite{Ablikim:2010aa} & $(0.31 \pm 0.16(\mathrm{stat}) \pm 0.14(\mathrm{sys}) \pm 0.03(\mathrm{para}))\%$ \\
  \hline
  $m_{K^+} \neq m_{K^0}$ & $M_{\pi\eta} \in [885,1085]\mathrm{~MeV}$ \\
  \hline
  in $T$ and $G$         & $0.26~\%$ \\
  only in $T$            & $0.19~\%$ \\
  only in $G$            & $0.05~\%$ \\
  \hline
  $m_{K^+} \neq m_{K^0}$ & $M_{\pi\eta} \in [m_\pi +m_\eta,1200\mathrm{~MeV}]$  \\
                         \hline
  in $T$ and $G$         & $0.17~\%$ \\
  only in $T$            & $0.12~\%$ \\
  only in $G$            & $0.03~\%$ \\
  \hline
\end{tabular}
\end{table}
\end{center}

We can see that we get a good agreement with the experimental measurements of BESIII \cite{Ablikim:2010aa} only when we introduce isospin breaking inside the $T$ matrix and also in the first external rescattering loops ($G_{K^{+} K^{-}}$ and $G_{K^{0}\bar{K}^{0}}$). The case where we use different kaon masses only inside the $T$ matrix is still within the range of the experimental errors, but the one where we consider them only in $G_{K^{+} K^{-}}$ and $G_{K^{0}\bar{K}^{0}}$ is already
outside the range of the experimental errors if they are summed in quadrature,
what shows the importance of considering both effects simultaneously. Besides, when we integrate in the more appropriate range of the $a_0(980)$ resonance, from 885 MeV to 1085 MeV as in Ref.~\cite{Ablikim:2006vm}, the results are closer to experiment.

\section{Conclusions}

We have shown in the present work that it is possible to use the same theoretical model previously developed to study the $\chi_{c1} \to \eta \pi^+ \pi^-$ reaction \cite{Liang:2016hmr}, recently measured by the BESIII Collaboration \cite{Kornicer:2016axs}, and further explored in the predictions for the $\eta_{c} \to \eta \pi^+ \pi^-$ reaction \cite{Debastiani:2016ayp}, to study the isospin breaking in the decays $\chi_{c1} \to \pi^0 \pi^+ \pi^-$ and $\chi_{c1} \to \pi^0 \pi^0 \eta$ and its relation to the $a_0(980) - f_0(980)$ mixing, which was also measured by the BESIII Collaboration \cite{Ablikim:2010aa}.

We assumed that the $\chi_{c1}$ behaves as an $SU(3)$ scalar to find the weight in which trios of pseudoscalars are created, followed by the final state interaction of pairs of mesons to describe how the $a_0(980)$ and $f_0(980)$ are dynamically generated, using the chiral unitary approach in coupled channels. The isospin violation was introduced through the use of different masses for the charged and neutral kaons, either in the propagators of the pairs of mesons created in the $\chi_{c1}$ decay, as in the propagators inside the $T$ matrix constructed through the unitarization of the scattering and transition amplitudes of pairs of pseudoscalar mesons.

We have analysed the contribution of each term and found that violating isospin inside the $T$ matrix makes the $\pi^0\eta \to \pi^+\pi^-$ contribution nonzero, which gives an important contribution to the total amplitude. We also investigated the importance of the isospin breaking from the $K\bar{K}$ term, and found that even tough the most important contribution comes from the first rescattering loops, violating isospin inside the $T$ matrix enhances this effect significantly. Also, in the total amplitude the most important effect is the isospin breaking inside the $T$ matrix, due to the constructive sum of $\pi^0\eta \to \pi^+\pi^-$ and $K\bar{K} \to \pi^+ \pi^-$, which is essential to get a good agreement with the experimental measurement of the mixing \cite{Ablikim:2010aa}.

\section*{Acknowledgments}

We would like to thank E.~Oset for suggesting the topic and for the fruitful discussions.

V.~R.~Debastiani wishes to acknowledge the support from the Programa Santiago Grisolia of Generalitat Valenciana (Exp.~GRISOLIA/2015/005), from the Institute of Modern Physics (Lanzhou) and Institute of Theoretical Physics (Beijing) of Chinese Academy of Sciences, and the interesting discussions with Ju-Jun~Xie, Feng-Kun~Guo and Bing-Song~Zou.


\bibliographystyle{plain}

\end{document}